\documentclass[aps,rmp,groupedaddress,longbibliography]{revtex4-1}

\usepackage{graphicx,color}
\usepackage{epsfig}
\usepackage{dcolumn}
\usepackage{amsmath,amssymb,bm}
\usepackage{hyperref,url}
\usepackage[noabbrev]{cleveref}
\usepackage{CJK}
\usepackage{lineno}
\usepackage[english]{babel}
\usepackage{subeqnarray}
\usepackage{mathptmx} 

\catcode`\@=11
\def\gtsim{\mathrel{\vcenter{\m@th\offinterlineskip
\hbox{$\hfill>\hfill$}\kern.5ex\hbox{$\hfill\sim\hfill$}}}}
\catcode`\@=12

\catcode`\@=11
\def\ltsim{\mathrel{\vcenter{\m@th\offinterlineskip
\hbox{$\hfill<\hfill$}\kern.5ex\hbox{$\hfill\sim\hfill$}}}}
\catcode`\@=12
\catcode`\@=12

\newcommand\Web{\mbox{We}}  

\begin{document}
\preprint{Submitted to Phys. Fluids}

\title{Transition from bubbling to jetting in a co-axial air-water jet}

\author{A. Sevilla, J. M. Gordillo and C. Mart\'{\i}nez-Baz\'an}
\affiliation{\'Area de Mec\'anica de Fluidos, Departamento de Ingenier\'{\i}a T\'ermica y de Fluidos, Universidad Carlos III de Madrid, 28911 Legan\'es, Spain}


\begin{abstract}
In this Brief Communication we study experimentally the flow regimes that appear in co-axial air-water jets discharging into a stagnant air atmosphere and we propose a simple explanation for their occurrence based on linear, local, spatiotemporal stability theory. In addition to the existence of a periodic bubbling regime for low enough values of the water-to-air velocity ratio, $u=u_w/u_a$, our experiments revealed the presence of a jetting regime for velocity ratios higher than a critical one, $u_c$. In the bubbling regime, bubbles form periodically from the tip of an air ligament whose length increases with $u$. However, when $u> u_c$ a long, slender gas jet is observed inside the core of the liquid coflow. Since in the jetting regime the downstream variation of the flow field is slow, we performed a local, linear spatiotemporal stability analysis with  uniform velocity profiles to model the flow field of the air-water jet. Similar to the transition from dripping to jetting in capillary liquid jets, the analysis shows that the change from the bubbling to the jetting regime can be understood in terms of the transition from an absolute to a convective instability.
\end{abstract}

\maketitle

A common mechanism to control the generation of gas bubbles is the use of a liquid coflow surrounding the gas injection needle, a configuration that produces  bubbles smaller than in the case without coflow\cite{chuang70,oguz93}. In this context, one of the most relevant parameters controlling the bubble size, or the bubble formation frequency, is the velocity ratio between the liquid coflow and the gas stream at the exit of the injection needle. For sufficiently small values of the liquid-to-gas velocity ratio, $u=u_w/u_a$, the formation of bubbles is a periodic process characterized by the non-linear growth and collapse of bubbles inside the liquid jet. However, in this brief communication we report experimental evidence of the transition from the aforementioned periodic \emph{bubbling regime} to an aperiodic \emph{jetting regime}, characterized by the formation of a long ligament of air inside the liquid jet, which occurs when the velocity ratio becomes larger than a critical value, $u_c$. This phenomenon is similar to the transition from dripping to jetting in free liquid jets\cite{ledizes97,clanet99}, where the formation of a jet from the nozzle is only possible for values of the Weber number higher than a critical one. Moreover, through the use of the concepts of locally convective and absolute instabilities, applied to a simplified flow model, we show that the jetting phenomenon is related to a convective instability, while the bubbling regime is the consequence of a transition to an absolute instability. Similar flow configurations have been previously studied from different point of views \cite{hertz1983,kendall86,chauhan1996,chauhan2000}. However these works focused mainly on the production of liquid shells in compound jets where the diameter of the outer jet is nearly the same as that of the inner one. In the present study, the air-water co-axial jet discharges into a stagnant air atmosphere~\citep{sevilla2005} avoiding, therefore, any effect of the outer shear layer~\citep{sevilla02}. Furthermore, unlike in the compound jet configuration studied by Kendall \cite{kendall86}, Chauhan et al. \cite{chauhan1996, chauhan2000} and others, we consider the diameter of the water jet to be much larger than that of the air jet and, consequently, the Rayleigh instability developing at the outer liquid interface does not affect the evolution of the inner, air flow.

\begin{figure}[h!]
\begin{center}
\includegraphics[width=0.3\textwidth]{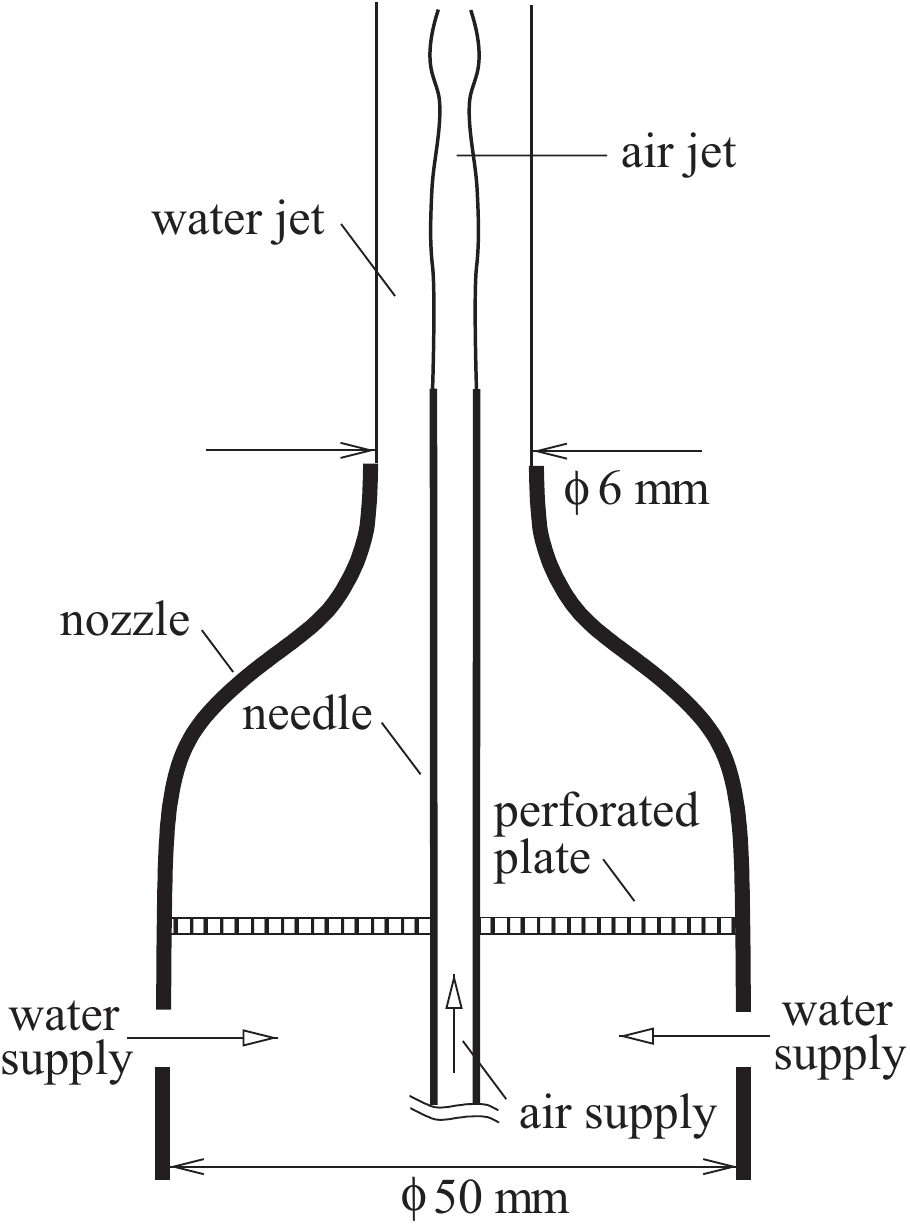}
\caption{Sketch of the experimental set-up} \label{fig1}
\end{center}
\end{figure}

The experiments were performed in the facility sketched in Fig. \ref{fig1}, where a jet of water discharges upwards into a still air atmosphere through a nozzle of radius $r_w=3\,\mbox{mm}$. An air stream was injected at the centerline of the water jet with a stainless steel hypodermic needle. The geometrical characteristics of  the different needles employed, hereafter named needles I, II and III, are summarized in table \ref{table1} where $r_i$, $r_o$ are  the inner and outer radii respectively. In the experiments presented here the water velocity was modified  from $u_w=1.85 \, {\rm m/s}$ to $u_w =9.65 \, {\rm m/s}$, and the air  velocity, controlled with a pressure regulator and a high precision valve, was varied from $u_a=2.7 \, {\rm m/s}$ to $u_a= 58.5 \, {\rm m/s}$. Measurements were performed by uniformly illuminating the measuring region with a white bulb light, and by recording images of the air-water co-axial jet with a Kodak Motion Corder Analyzer high speed camera. In the present work the images recorded were taken at a rate of $2000$ frames per second with a resolution of $256 \times 120$. The relevant control parameters of the present configuration are the Weber number, defined as $\Web=\rho_au_a^2r_o/\sigma$ where $\rho_a$ denotes the air density and $\sigma$ the surface tension coefficient, which in the experiments reported here varied between $\Web \approx 0.08$ and $\Web \approx 27$, the water-to-air velocity ratio, $u=u_w/u_a$, with $0<u\lesssim 1.25$, and the geometrical parameters, namely the water-to-air diameter ratio, $r_w/r_o$, and the ratio of the outer to inner needle radius, $r_o/r_i$. However, the Weber number of the water stream, $\rm{\Web_w}=\rho_w \, u_w^2 \, r_w/\sigma \gg 1$, was always sufficiently large to neglect capillary effect at the outer water-air interface.

\begin{table}
\centering
\begin{tabular}{c c c c}
\hline
   Needle & $r_i(\mbox{mm})$ & $r_o(\mbox{mm})$ & $r_w/r_o$
   \\\hline
   I & 0.597 & 0.8255 & 3.6 \\
   II & 0.419 & 0.635 & 4.7 \\
   III & 0.292 & 0.451 & 6.6 \\
\hline
\end{tabular}
\caption{Geometrical properties of the three different needles used in the experiments. Here $r_i$ and $r_o$ are the inner and outer radius of the air injection needle respectively and $r_w$ is the radius of the water jet.} \label{table1}
\end{table}

\begin{figure}[h!]
\begin{center}
\includegraphics[width=0.7\textwidth]{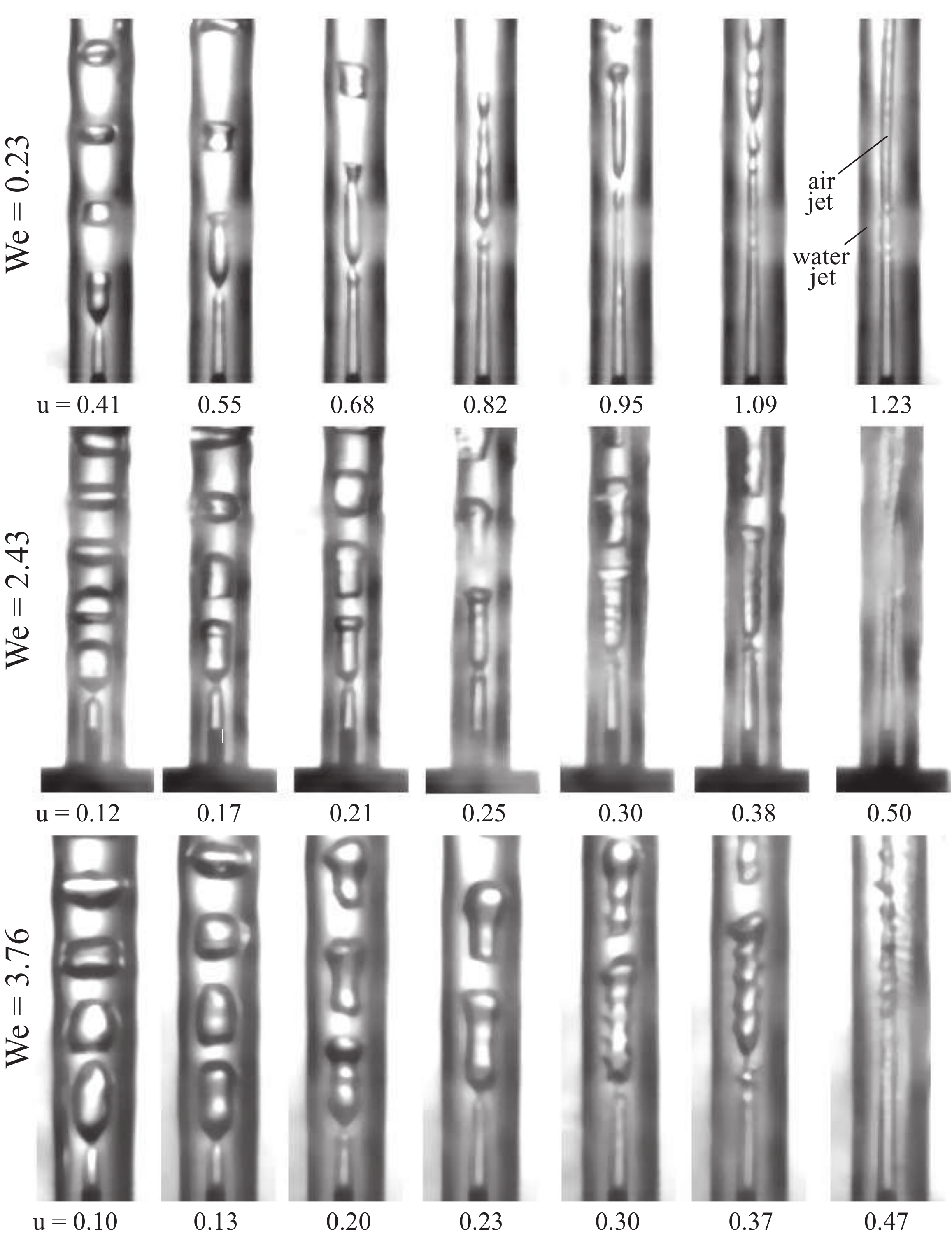}
\caption{Photographs showing experimental evidence of the transition from bubbling to jetting regimes for three different values of $\Web$ and increasing values of $u$. The top panels have $\Web$=0.23 while the central panels have $\Web$=2.43 and the bottom panels $\Web$=3.76. The water-to-air velocity ratio is indicated in each image.} \label{fig2}
\end{center}
\end{figure}

Figure \ref{fig2} shows some snapshots taken at the instant of bubble detachment, which correspond to experiments performed with needle II for three different values of the Weber number, namely $\Web=0.23$, 2.43 and 3.76 respectively, and increasing values of $u$.

Note that, for a fixed value of $\Web$, the length of the air ligament from which bubbles detach periodically increases with $u$. Moreover, when $u$ reaches a critical value, $u_c$, a transition from the periodic \emph{bubbling regime} to the \emph{jetting regime} takes place, as indicated in the last photograph of the first row in Fig. \ref{fig2}. It can also be observed that the value of the critical velocity ratio depends on the Weber number, $u_c=u_c(\Web)$. Notice that for the first row of pictures in Fig. \ref{fig2}, where $\Web=0.23$, theregime is still bubbling for a velocity ratio $u=1.09$, while a jetting phenomenon can be seen for $u=1.23$, so we can conclude that $1.09<u_c(\Web=0.23)<1.23$. Similarly, for the second and third rows we find $0.38<u_c(\Web=2.43)<0.50$ and $0.37<u_c(\Web=3.76)<0.47$.

As can be inferred from Fig. \ref{fig2}, in the jetting regime the air-water flow is slowly divergent and the use of parallel linear instability analysis is justified on physical grounds. Keeping this idea in mind, we studied the parallel, spatiotemporal stability of the simplest basic flow which retains the main physical mechanisms to model the dynamics of the coflowing air water jet. Thus, both the gas and the liquid streams were approximated by parallel coaxial jets, with radii $r_a$, $r_w$, uniform velocities $u_a$, $u_w$ and densities $\rho_a$, $\rho_w$ respectively. Viscous effects were neglected in the analysis, since they are known to be small corrections for the high values of the gas and liquid Reynolds numbers considered here. Although calculations were also performed  using a more realistic, parabolic velocity profile for the air steam, the results obtained were very similar to the case of uniform velocity profiles, as far as the absolute or convective character of the flow was concerned. Here, cylindrical coordinates will be denoted ($x,\,r,\,\theta$), and normal modes of the pressure disturbance will be written as $p\,e^{i(kx+m\theta-\omega t)}$, where $p=p(r)$ is the radial eigenfunction, and $k,\,m,\,\omega$ are the axial wavenumber, azimuthal number and angular frequency respectively. Since we are considering a piecewise uniform basic velocity profile, the pressure eigenfunctions take the following form,
\begin{subeqnarray}\label{eigen}
     p_a & = & A\mbox{I}_m(kr)+B\mbox{K}_m(kr),\;\;0\leq r\leq r_a,\\
     p_w & = & C\mbox{I}_m(kr)+D\mbox{K}_m(kr),\;\;r_a\leq r\leq r_w,
\end{subeqnarray}
where $\mbox{I}_m$ and $\mbox{K}_m$ are the first and second kind modified Bessel functions of order $m$ respectively. Equations (\ref{eigen}a) and (\ref{eigen}b) are subjected to the following boundary conditions at $r=r_a$ and $r=r_w$,
\begin{subeqnarray}\label{boundary}
     r&=&0:\;p_a\neq\infty,\\
     r&=&r_a:\;p_a-p_w = 
\frac{\sigma(k^2r_a^2+m^2-1)}{\rho_ar_a^2(ku_a-\omega)^2}\frac{dp_a}{dr},\\
r&=&r_a:\;\frac{1}{\rho_a(ku_a-\omega)^2}\frac{dp_a}{dr}=\frac{1}{\rho_w(ku_w-\omega)^2}\frac{dp_w}{dr},\\
     r&=&r_w:\;p_w=0,
\end{subeqnarray}
where equation (\ref{boundary}a) is  the regularity condition at the axis, and equations (\ref{boundary}b)--(\ref{boundary}d) are the linearized expressions of the pressure jump condition across the interface, the kinematic condition at the interface and the condition of ambient pressure at the outer water-air interface respectively. Equation (\ref{boundary}d) was applied since $\rm{\Web_w}=\rho_w \, u_w^2 \, r_w/\sigma \gg 1$ and, therefore, the pressure at the outer liquid interface remained unperturbed. Introducing equations (\ref{eigen}a)--(\ref{eigen}b) into (\ref{boundary}a)--(\ref{boundary}d), and imposing the existence of non-trivial solutions of the problem, we obtain the following dispersion relation~\cite{chauhan1996,ledizes97,lishen98,sevilla02},
\begin{equation}\label{axiladr}
\left[1-\frac{\alpha\left(\alpha^2+m^2-1\right)}{\Web\left(\alpha-\beta\right)^2}
\frac{\mbox{I}'_m\left(\alpha\right)}{\mbox{I}_m\left(\alpha\right)}\right]
\left(\frac{\mbox{K}'_m\left(\alpha\right)}{\mbox{I}'_m\left(\alpha\right)}-
\frac{\mbox{K}_m\left(\alpha
\,r_w/r_a\right)}{\mbox{I}_m\left(\alpha\,r_w/r_a\right)}\right)=\frac{\rho_w}{\rho_a}\left(\frac{\alpha
u-\beta}{\alpha-\beta}\right)^{2}\left(\frac{\mbox{K}_m\left(\alpha\right)}{\mbox{I}_m\left(\alpha\right)}-
\frac{\mbox{K}_m\left(\alpha\,r_w/r_a\right)}{\mbox{I}_m\left(\alpha\,r_w/r_a\right)}\right)\,,
\end{equation}
where primes denotes the derivative of the corresponding Bessel function and $\alpha=kr_a$ and $\beta=\omega r_a/u_a$ are the dimensionless wavenumber and frequency respectively. The convective or absolute nature of the instability of the model flow can be studied by examining the temporal growth rate of the local modes with zero group velocity, $d\beta/d\alpha=0$ (see Huerre\cite{huerre00} and references therein). Previous works have already shown that, for the case of liquid-gas jets, the dispersion relation equivalent to (\ref{axiladr}) supports absolute instability when the liquid Weber number becomes smaller than a critical value, which depends on the velocity ratio\cite{ledizes97,lishen98}. Similarly, here we show that, for the air-water jet under study ($\rho_a/\rho_w\approx 1.2\times 10^{-3}$), equation (\ref{axiladr}) predicts convective instability in the axisymmetric mode ($m=0$) when the velocity ratio is larger than a critical value which depends on the gas Weber number and on the diameter ratio.

\begin{figure}[h!]
\begin{center}
\includegraphics[width=0.6\textwidth]{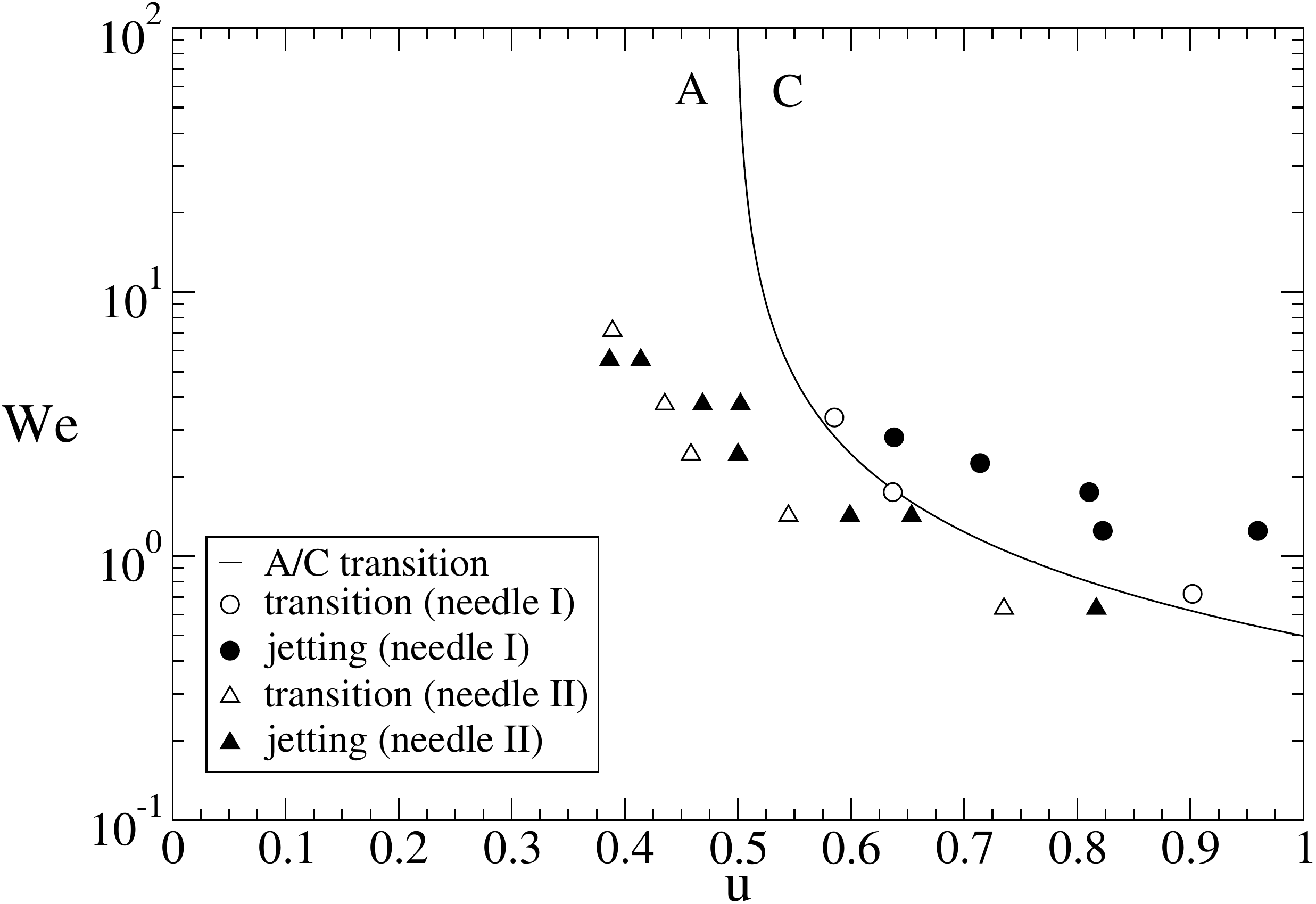}
\caption{Comparison between the absolute/convective transition curve in the $\Web$\,--\,$u$ plane (solid line) with the experimental bubbling/jetting transition diagram for needles I (circles) and II (triangles). Open symbols indicate situations where a transition from bubbling to jetting was observed, while solid symbols indicate jetting regimes.} \label{fig3}
\end{center}
\end{figure}

Figure \ref{fig3} shows the result of a systematic experimental study of the transition from \emph{bubbling} to \emph{jetting}, together with the predictions given by the dispersion relation \eqref{axiladr}. We calculated the theoretical absolute/convective, A/C, transition curve in the $\Web$--$u$ plane for the values of $r_w/r_a$ corresponding to our experimental conditions, namely $r_w/r_a$=3.6, 4.7 and 6.6. However, no appreciable difference was found among these three cases, and the solid line in Fig. \ref{fig3} corresponds to the prediction given by the stability analysis in the range of diameter ratios $3.6<r_w/r_a<6.6$. Notice that convective instability is only possible for a sufficiently large value of the velocity ratio, which decreases as the Weber number increases, reaching an asymptotic value of $u_c\approx 0.5$ as $\Web\rightarrow\infty$. Figure $\ref{fig3}$ also shows our experimental observations, including points which corresponded to situations where the flow behavior changed randomly from the \emph{bubbling regime} to the \emph{jetting regime} and viceversa, here referred to as \emph{transition points}. Jetting and transition events are plotted in Fig. \ref{fig3} as filled and open symbols, with needles I and II represented by circles and triangles, respectively. The absence of transition or jetting phenomena for needle III was due to experimental limitations: since the diameter of needle III was very small, the velocity of the gas stream was very large even for small gas flow rates and, correspondingly, the maximum velocity ratio obtained in this case was not sufficient to observe the \emph{jetting regime}. Although there were many bubbling points in the diagram, we decided not to include them in Fig. \ref{fig3} for clarity. Let us comment on the main features of the transition diagram. First at all, notice that the transition to \emph{jetting regime} is observed for a high enough value of the velocity ratio, $u_c$, which increases as the Weber number decreases. On the other hand, a jetting  behavior was never observed for velocity ratios smaller than $\sim 0.4$, regardless of the value of the Weber number. Notice that the agreement is excellent for the case of needle I where it may be observed that the experimental jetting events (solid circles) are inside the convectively unstable region of the diagram, while transitional events (hollow circles) fall onto the absolute/convective curve and bubbling events (not shown for clarity) fall within the absolutely unstable region. However, the experiments carried out with needle II slightly deviate from the results of the instability analysis, obtaining a few jetting events (solid triangles) within the absolutely unstable region of the transition diagram. The theoretical transition curve seems to overestimate the critical velocity ratio observed in the experiments. Nevertheless, if we followed the evolution of the transition points obtained with needle II, represented with hollow triangles in Fig. \ref{fig3}, we would appreciate the same trend as in the case of needle I, indicating the existence of an asymptotic critical velocity ratio for large values of the Weber number. The difference between the experimental results obtained with needles I and II might be attributed to a misestimation of the air velocity $u_a$, calculated as $u_a=Q_a/(\pi r_i^2)$ where $Q_a$ is the air flow rate experimentally measured with a flow meter. A close look at the air ligaments formed during the experiments performed with needle II showed that their diameters were in fact slightly larger than $r_i$, indicating that  the experimental value of $u$ might have been underestimated for the points corresponding to needle II in Fig. \ref{fig3}. Nevertheless, the results of the stability analysis are in good agreement with the experiments demonstrating that the appearance of the jetting regime may be attributed to the convectively unstable nature of the flow. In this context, we would like to emphasize that the periodic bubbling regime is an intrinsically non-linear phenomenon which is not expected to be described by a linear stability mechanism. Thus, the linear analysis just shows that the formation of a jet is only possible when the flow is globally stable. 

\begin{acknowledgments}
This research was supported by the Spanish MCyT under Project \# DPI2002-04550-C07.
\end{acknowledgments}


%

\end{document}